\documentclass[journal,twocolumn,web]{ieeecolor}
\usepackage{generic}
\usepackage{amsmath,amssymb,amsfonts}
\usepackage{algorithmic}
\usepackage{graphicx}
\usepackage{textcomp}

\usepackage{array}
\usepackage{adjustbox}
\usepackage{hyperref}
\usepackage{booktabs}
\usepackage{subcaption}
\usepackage[sorting=none, backend=biber, style=ieee, url=false, doi=false, maxbibnames=3]{biblatex}
\newcommand{\supplementarysection}{%
  \setcounter{figure}{0}% Reset figure counter
  \let\oldthefigure\thefigure% Capture figure numbering scheme
  \renewcommand{\thefigure}{S\oldthefigure}% Prefix figure number with S
   \setcounter{table}{0}% Reset table counter
  \let\oldthetable\thetable% Capture table numbering scheme
  \renewcommand{\thetable}{S\oldthetable}% Prefix table number with S
  \section{Supplementary section}% Set supplementary section
}

\def\BibTeX{{\rm B\kern-.05em{\sc i\kern-.025em b}\kern-.08em
    T\kern-.1667em\lower.7ex\hbox{E}\kern-.125emX}}
\begin{document}
\title{Machine Learning for Ranking f-wave Extraction Methods in Single-Lead ECGs}
\author{Noam Ben-Moshe, Shany Biton, Kenta Tsutsui, Mahmoud Suleiman, Leif S{\"{o}}rnmo  \textit{Fellow, IEEE}, and Joachim A. Behar, \textit{Senior Member, IEEE}
\thanks{Manuscript submitted on \today. The research was supported for NBM, SB, and JB by a grant (3-17550) from the Ministry of Science \& Technology, Israel \& Ministry of Europe and Foreign Affairs (MEAE) and the Ministry of Higher Education, Research and Innovation (MESRI) of France. SB, NBM, MS, and JAB acknowledge the support of the Technion-Rambam Initiative in Artificial Intelligence in Medicine, Hittman: Technion EVPR Fund: Hittman Family Fund and Israel PBC-VATAT and by the Technion Center for Machine Learning and Intelligent Systems (MLIS). (Corresponding author: jbehar@technion.ac.il)}
\thanks{N. Ben-Moshe, is with the Faculty of Computer Science and Faculty of Bio-Medical, Technion-IIT, Haifa, Israel.}
\thanks{S. Biton, and J. A. Behar are with the Faculty of Bio-Medical, Technion-IIT, Haifa, Israel.}
\thanks{M. Suleiman is with the Department of Cardiology, Rambam Medical Center and Technion The Ruth and Bruce Rappaport Faculty of Medicine, Haifa, Israel.}
\thanks{L. S\"ornmo is with the Department of Biomedical Engineering, Lund University, Lund, Sweden.}
\thanks{K. Tsutsui is with the Department of Cardiovascular Medicine, Faculty of Medicine, Saitama Medical University International Medical Center, Saitama, Japan}
}
\maketitle
% \nocite{*}
% Shany Biton, Mahmoud Suleiman, Leif S\"ornmo and  Joachim A. Behar
% may add some middle authors; Kenta.

\begin{abstract}
%Put a space between authors' initials. 
%The abstract should include three or four different keywords or phrases, as this will help readers to find it. 
%The abstract must be between 150--250 words.

\textbf{Introduction}: The presence of fibrillatory waves (f-waves) is important in the diagnosis of atrial fibrillation (AF), which has motivated the development of methods for f-wave extraction. We propose a novel approach to benchmarking  methods designed for single-lead ECG analysis, building on the hypothesis that better-performing AF classification using features computed from the extracted f-waves implies better-performing extraction. The approach is well-suited for processing large Holter data sets annotated with respect to the presence of AF. \textbf{Methods}: Three data sets with a total of 300 two- or three-lead Holter recordings, performed in the USA, Israel and Japan, were used as well as a simulated single-lead data set. Four existing extraction methods based on either average beat subtraction or principal component analysis (PCA) were evaluated. A random forest classifier was used for window-based AF classification. Performance was measured by the area under the receiver operating characteristic (AUROC). \textbf{Results}: The best performance was found for PCA-based extraction, resulting in AUROCs in the ranges 0.77--0.83, 0.62--0.78, and 0.87--0.89 for the data sets from USA, Israel, and Japan, respectively, when analyzed across leads; the AUROC of the simulated single-lead, noisy data set was 0.98. \textbf{Conclusions}: This study provides a novel approach to evaluating the performance of f-wave extraction methods, offering the advantage of not using ground truth f-waves for evaluation, thus being able to leverage real data sets for evaluation. The code is open source (following publication).
\end{abstract}


\begin{figure}[h]
\centering
\includegraphics[page=1,width=1\columnwidth]{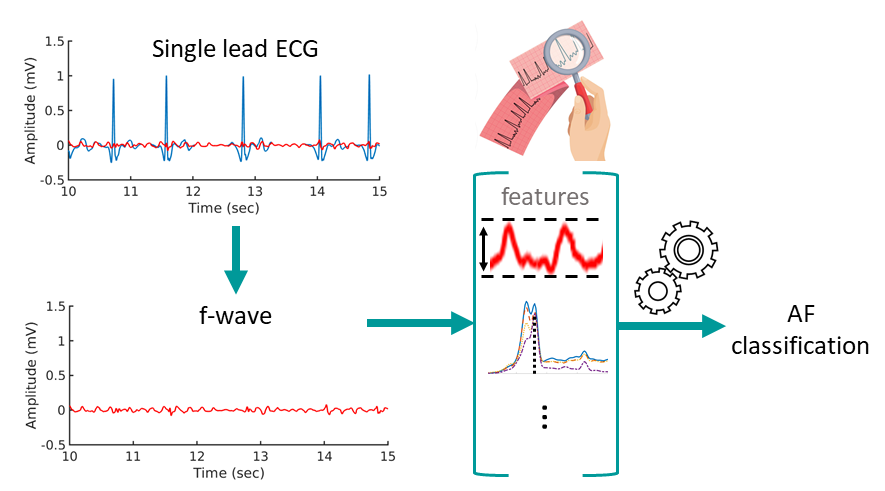}

\caption{An ECG is processed to estimate the f-wave using an extraction method. A set of f-wave features is used to train a machine learning model for AF classification. The performance of AF classification serves as a measure to rank the performance of different extraction methods. Part of the figure is adapted from Biorender. }
\label{graphical_abstruct}
\end{figure}

\section{Introduction}
Fibrillatory waves (f-waves) are present in the ECG recorded from patients with atrial fibrillation (AF), representing the rapid and disorganized depolarization of the atria. Diagnosis of AF requires irregular RR~intervals, absence of P-waves, and presence of f-waves. Therefore, f-wave information is important for diagnosis in order to distinguish AF from other arrhythmias such as atrial flutter (AFL) and multifocal atrial tachycardia. 
% \cite{AFLLearnTheHeart}

Thanks to its simplicity, average beat subtraction (ABS) remains the most widely used f-wave extraction method in clinically oriented studies, building on the observation that atrial activity is decoupled from ventricular activity~\cite{LSornmo2018ExtractionWaves}. However, ABS suffers from several limitations due to the assumptions of the underlying model of stable QRS morphology and fixed noise level. Therefore, to offer better performance, several variants of ABS have been proposed \cite{Dai2013AtrialMethods,Bataillou1995WeightedWeights, Stridh2001SpatiotemporalFibrillation, Lemay2007CancellationMethods}, as well as blind source separation-based methods, including principal analysis~\cite{Castells2005EstimationConcepts}, periodic~\cite{Mihandoost2022AExtraction}, and independent component analysis~\cite{Rieta2004AtrialSeparation}. Most of these methods assume that a multi-lead ECG is available.

The performance of f-wave extraction methods has been benchmarked by means of real as well as simulated ECG signals~\cite{LSornmo2018ExtractionWaves}. Depending on the type of signal, different local measures have been used to quantify performance. For real ECGs, indirect measures have been used to quantify the large changes in f-wave amplitude which imply greater QRS residuals and poor f-wave extraction \cite{Alcaraz2008AdaptiveElectrocardiograms, Lee2012EventElectrocardiograms, Malik2017Single-leadGeometry}. A disadvantage with such measures is their blindness to the spectral properties of the extracted signal, which are significant since the extracted signal typically exhibits a dominant frequency in the interval $[4,12]$~Hz. 

For simulated ECGs, sample-by-sample measures quantifying the error between extracted and true f-wave signal has been used, e.g., in terms of the mean square error or the cross-correlation coefficient~\cite{Lemay2007CancellationMethods,Roonizi2017AnFibrillation,Mateo2013RadialFibrillation}. The use of simulated ECGs offers the advantage of having access to the ground truth, but, on the other hand, it comes with the disadvantage of not fully accounting for the physiological variability which exists across patients, neither for various types of real life noise and artifacts.

The standard 12-lead ECG continues to be used in clinical practice to diagnose AF, where lead V$_1$ offers the largest f-waves due to its proximity to the atria~\cite{APetrenas2018LeadDevice}. However, single-lead ECGs are becoming increasingly more common thanks to the development of patches and smartwatches for remote health monitoring and screening~\cite{Steinhubl2018EffectTrial, Barrett2014ComparisonMonitoring}. 
Though single-lead ECGs cannot be used to confidently diagnose certain heart diseases, recent research has nonetheless shown that single-lead ECGs have a high clinical potential for detecting AF~\cite{Kaasenbrood2016YieldVaccination, Svennberg2017SafeFibrillation, Chocron2021RemoteNetwork,BenMoshe2022arnetECG, Biton2023GeneralizableSexes}. Since most research on f-wave extraction have assumed that multi-lead ECGs are available, there is an urgent need to develop methods for robust handling of single-lead ECGs.

The primary objective of the present study is to propose an approach to ranking of f-wave extraction methods designed for single-lead ECGs, whether real or simulated in nature. The approach builds on the hypothesis that better-performing AF classification using a set of features computed from the extracted f-waves implies better-performing extraction. The features are used as input to a machine learning model applied to successive, non-overlapping windows. The secondary objective is to investigate how the best-performing extraction method depends on lead position, age, and sex.

The paper is organized as follows. Section~\ref{secMaterials} describes the three real data sets consisting of Holter ECG recordings and the simulated data set. Section~\ref{secMethods} describes the f-wave extraction methods subject to evaluation and the machine learning approach taken to AF classification. Section~\ref{secResults} presents the performance on real and simulated data, followed by a discussion in Sec.~\ref{secDiscussion} of the results and finally the conclusions in Sec. ~\ref{secConclusion}.

% A possible approach to assess f-wave extraction performance is to employ global measures. Accordingly, we hypothesized that better extraction of the f-wave meant better AF/non-AF classification using features engineered from the extracted f-wave. For that purpose features of the extracted f-wave are engineered and used as input to a machine learning model which task is to classy a given window as AF or non-AF.

\begin{table}
\caption{Description of data sets. Age is presented as median and interquartile range (Q1--Q3).}
\label{table:DemographicDescription}
\centering
\begin{tabular}{lccl}
\toprule
& \textbf{UVAF}& \textbf{SHDB}& \textbf{RBDB}\\
       \midrule
    Origin & USA  & Japan & Israel  \\  [0.5ex]
    Patients, $n$ & 100 & 100 &100 \\  [0.5ex]
    Age (yrs) & 69 (59--76) & 70 (62--75) & 70 (57--78) \\  [0.5ex]
    Female, $n_f$ & 50 & 45 & 50 \\  [0.5ex]
\bottomrule
\end{tabular}
\end{table}

% If you are submitting your paper to a colorized journal, you can use
% the following two lines at the start of the article to ensure its
% appearance resembles the final copy:

\smallskip\noindent
\begin{small}
\begin{tabular}{l}
% \verb+\+\texttt{documentclass[journal,twoside,web]\{ieeecolor\}}\\
% \verb+\+\texttt{usepackage\{\textit{Journal\_Name}\}}
\end{tabular}
\end{small}

%%%%%%%%%%%%%%%%%%%%%%%%%%%%%%%%%%
\section{Materials}
\label{secMaterials}
\subsection{Real data sets} 
Three different data sets were used: the University of Virginia Atrial Fibrillation data set (UVAF), USA \cite{Chugh2014WorldwideStudy, Moss2014LocalImplications}, the Rambam Hospital Holter clinic data set (RBDB), Israel~\cite{Biton2023GeneralizableSexes}, and the Saitama Hospital data set (SHDB), Japan~\cite{Biton2023GeneralizableSexes}. UVAF consists of three-lead Holter recordings for which no lead information is available. The original data set consists of 2,147 patients totaling 51,386 hours of continuous ECG recordings. As no patient reports were available, the diagnoses were inferred from AF episode annotation~\cite{Moss2014LocalImplications}. RBDB consists of medical reports and three-lead Holter recordings (leads CM5, CC5, and CM5R), except when battery life had to be saved to handle recordings exceeding 24~h, then instead resulting in two-lead recordings (CM5 and CC5). SHDB consists of medical reports and two-lead Holter ECG recordings (leads NASA and CC5). 

% In compliance with all relevant ethical regulations ethical approval for using the retrospective de-identified data was granted by the Saitama Medical University institutional ethics committee under IRB number 20–030 and the Rambam Health Care Campus institutional ethics committee under IRB: D-0402-21.

A subset of 100 recordings was used from each data set, similar to what was done in our previous work~\cite{Biton2023GeneralizableSexes}. Briefly, stratification was performed according to age, sex, and AF diagnosis. Based on the cardiology report of each data set, 80 recordings were chosen from patients with AF. Each recording was manually reviewed and annotated with respect to AF episodes by an expert cardiologist~\cite{Biton2023GeneralizableSexes}. Recordings were digitized at a sampling rate of 200 Hz. The median length of the recordings used in each data set was 24~h. 

%Each recording was divided into non-overlapping 1-min windows. The recording was divided to maximize the number of windows without mixed labels of AF and non-AF. 

Each recording was divided in 1-min non-overlapping windows. Windows with mixed rhythms, defined as AF and non-AF rhythms, were excluded. The following two criteria were used to determine whether a window was to be excluded due to low signal quality: 1.~the window contained too few QRS complexes (less than 10~QRS complexes), and 2.~the signal quality index bSQI~\cite{Behar2013ECGReduction, Li2008RobustSources} was below 0.8. In addition, windows annotated as AFL were excluded. The exclusion criteria were applied independently to each lead.  Overall, 21\% of all windows were excluded from the analysis. The number of windows in each lead and the number of excluded windows are listed in the supplement (Fig.~\ref{exclusion_of_windows}).

\subsection{Simulated data set}
% A simulated data set was generated using the model described in~\cite{APetrenas2017ElectrocardiogramEpisodes}. The ECGs were chosen to be noise-free in order to establish an upper bound on performance. A total of 500 12-lead ECGs were simulated, all with a duration of about 5~min, and lead V$_1$ was subject to analysis. Real components were used for simulation, i.e., ventricular rhythm, atrial activity (f- or P-waves), and QRST complexes were randomly chosen from different databases. The AF burden was chosen randomly from a uniform distribution defined by the interval $[0,1]$. The recording was divided into windows in the same way as the real data sets.

Two simulated data sets were generated using the model described in~\cite{APetrenas2017ElectrocardiogramEpisodes}. One data set was simulated without noise and the other with noise. The former set was simulated for the purpose of establishing an upper bound on performance, whereas the latter set was simulated as a mixture of baseline wander, muscle noise, and electrode movement artifacts, having a noise level of 100~$\mu$V (root mean square, RMS) which can be viewed as representative of Holter recordings. 

In each data set a total of 500 12-lead ECGs were simulated, all with a duration of about 5~min; lead V$_1$ was subject to analysis. Real components were used for simulation, i.e., ventricular rhythm, atrial activity (f- or P-waves), and QRST complexes were randomly chosen from different databases with real ECGs. The AF burden was chosen randomly from a uniform distribution defined by the interval $[0,1]$. Each simulated recording was divided into windows in the same way as the real data sets.

%-----------------------------------------------------------------
\section{Methods}
\label{secMethods}
\subsection{f-wave extraction}

The ECG were filtered using a zero-phase, second-order bandpass filter, with a passband of 0.67--100~Hz to remove baseline wander \cite{Kligfield2007RecommendationsElectrocardiogram} and high-frequency noise. Depending on data set, a notch filter at 50 or 60~Hz was used to remove powerline interference. Single-lead QRS detection was performed using the detector in~\cite{Pan1985AAlgorithm}.  

We evaluate four different methods proposed for f-wave extraction~\cite{LSornmo2018ExtractionWaves}, also suitable for the analysis of the fetal ECG in the presence of the maternal ECG~\cite{Behar2014CombiningData, Behar_2016}. The code implementation is provided in the open resource  \href{http://www.fecgsyn.com}{fecgsyn.com}~\cite{Behar2014CombiningData}. The basic ABS method computes the ensemble average of time-aligned cardiac cycles, serving as a QRST template which is subtracted from the original ECG~\cite{Slocum1992DiagnosisActivity}. The resulting residual signal contains f-waves but also QRST-related residuals and noise. Variants of ABS offer different degrees of adaptability of the QRST template with respect to amplitude scaling. Either the entire QRST template is scaled by a factor before subtraction~\cite{Beckers2005DeterminationLeads} or the QRS~complex and the T-wave of the template are scaled individually~\cite{Martens2007ARecordings}. These two methods are denoted ABS$_{\textrm{sc1}}$ and ABS$_{\textrm{sc2}}$, respectively. 

Using principal component analysis (PCA), the almost periodic characteristic of the ECG is explored for selective separation of the ventricular activity and the f-waves~\cite{Kanjilal1997FetalDecomposition}, see also~\cite{Castells2005EstimationConcepts}. This method performs eigenvector-based template subtraction (TS) and therefore denoted~TS$_{\textrm{PCA}}$. 

\begin{figure*}[h]
\centering
\includegraphics[page=1,width=1\textwidth]{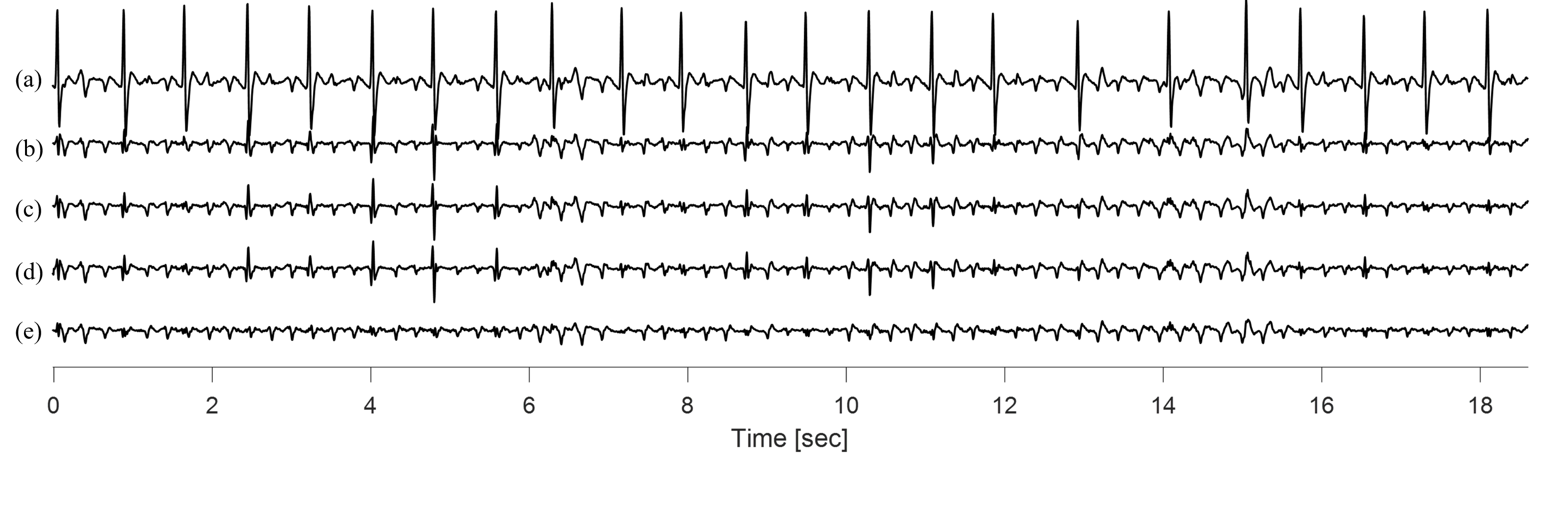}
\vspace{-1cm}
\caption{Illustration of f-wave extraction. (a) A single-lead ECG with AF and related f-waves extracted using (b)~ABS, (c)~ABS$_{\textrm{sc1}}$, (d)~ABS$_{\textrm{sc2}}$, and (e)~TS$_{\textrm{PCA}}$. In this example, TS$_{\textrm{PCA}}$ is associated with the smallest QRS-related residuals. }
\label{exampleFwave}
\end{figure*}

The extraction of f-waves is performed in every 1-min window and the resulting signal is denoted $d(n), n=1,\hdots,N$, where $N$ is the number of samples in the window. Examples of f-wave signals extracted using ABS, ABS$_{\textrm{sc1}}$, ABS$_{\textrm{sc2}}$, and ~TS$_{\textrm{PCA}}$ are presented in Fig.~\ref{exampleFwave}. 

%Matlab R2021a and Python 3.8 were used. 

%-----------------------------------------------------------------
\subsection{f-wave features} \label{subsection:feature_engineering}
\begin{figure}[t]
%\centering
\includegraphics[page=1,width=\columnwidth]{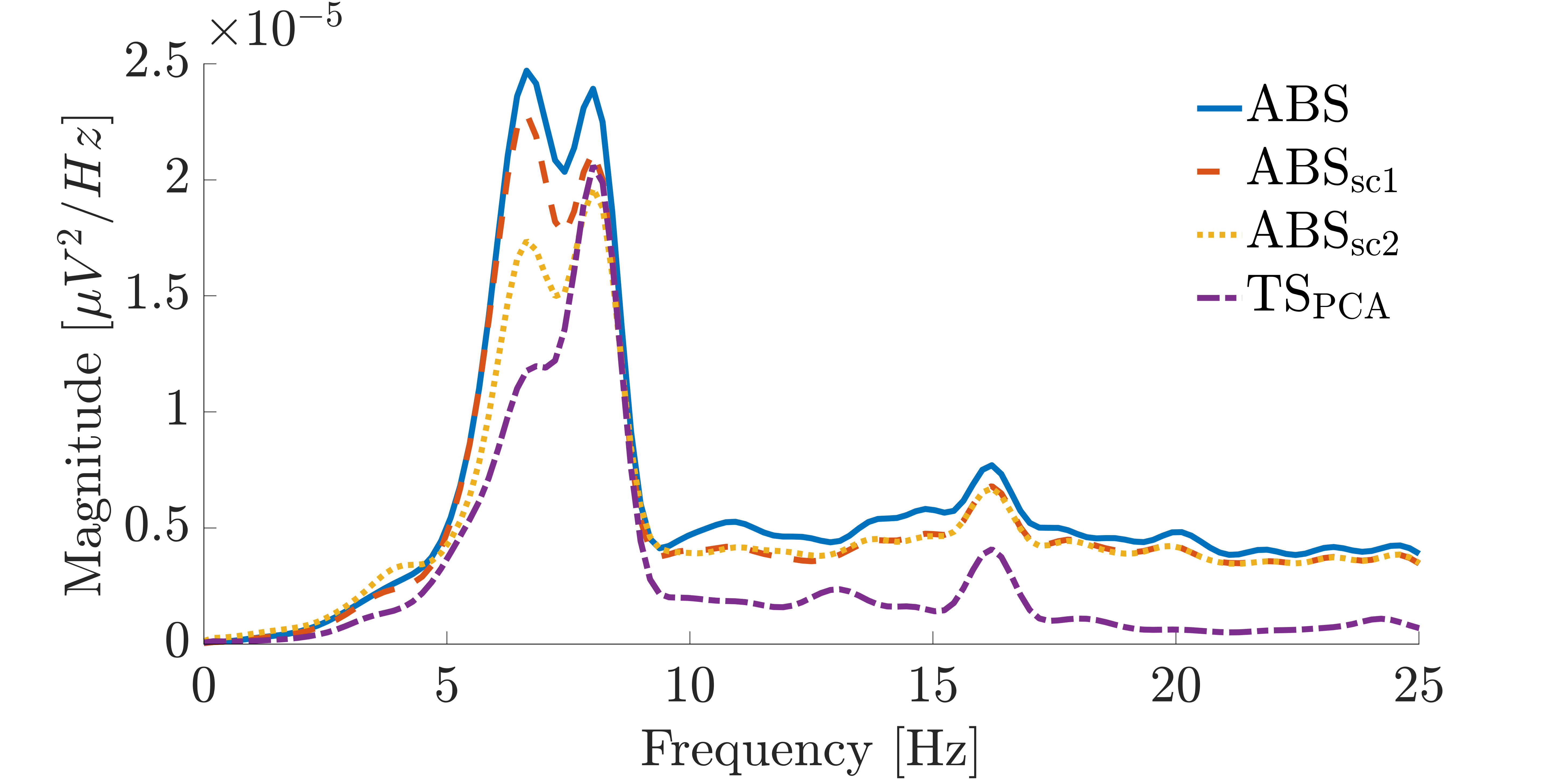}
\caption{Example of power spectrum for extracted f-wave signals obtained by the four different methods. }
\label{examplePS}
\end{figure}

A total of five features were computed, one defined in the time domain and four in the frequency domain. The \textit{overall peak-to-peak amplitude} of the extracted f-wave signal is defined by
\begin{equation} 
A_{pp} = \max_{n \in \Omega_f}(d(n)) - \min_{n \in \Omega_f}(d(n)),
\end{equation}
where $\Omega_f$ is defined by all samples of $d(n)$, excluding those which are inside the QRS interval, whose limits are set 90 ms before and 90ms after the detected R-peak.
 The power spectrum  of $d(n)$, denoted $P_d(\omega)$, was computed using Welch's method with a Hamming window and 50\% segment overlap, see Fig.~\ref{examplePS} for an example. The \textit{dominant atrial frequency} (DAF) is defined by the position of the largest peak in the interval [4,12]~Hz~\cite{LSornmo2018ExtractionWaves}, corresponding to the normalized frequencies $\omega_{l}$ and $\omega_{u}$, respectively,
\begin{equation} 
\omega_{\textrm{DAF}} = \arg \max_{\omega_{l} \leq \omega \leq \omega_{u}}(P_d(\omega)).
\end{equation}
Once $\omega_{\textrm{DAF}}$ is determined, the \textit{DAF magnitude} is given by
\begin{equation} 
P_{\textrm{DAF}} = P_d(\omega_{\textrm{DAF}}).
\end{equation}
The \textit{power inside the interval} [4,12]~Hz and the \textit{power outside the interval} are defined by 
\begin{align}
P_{i} &= \int_{\omega_{l}}^{\omega_{u}} P_d(\omega) \,d\omega, \\
P_{o} &= \int_{0}^{\omega_{l}} P(\omega) \,d\omega  + \int_{\omega_{u}}^{\pi} P(\omega) \,d\omega, 
\end{align}
respectively.%, which are used to define the \textit{spectral power ratio},
% \begin{equation}  
% R_{i/o} = \frac{P_{i}}{P_{o}}.
% \end{equation}

\subsection{Machine learning}
Based on the above five f-wave features, a random forest (RF) classifier was used to determine whether AF was present in 1-min windows. Classifier performance was measured by the area under the receiver operating characteristic (AUROC). Each of the four data sets were divided into a training set and a test set using a 80/20 split. For each training set and extraction method, a Bayesian search with a fivefold cross-validation was performed for hyperparameter tuning, using the search space given in Table~\ref{table:hyperparameter_space}. The test sets were sampled 100 times and the AUROC was computed for each of the sampled sets. For each extraction method and data set the median was reported. The confidence interval of the AUROC was evaluated similar to~\cite{Biton2021AtrialLearning}.

% , the Q1 and Q3 quartiles were 

\begin{table}[ht]
\caption{ Hyperparameter search space for grid search.}
\label{table:hyperparameter_space}
\centering
\begin{tabular}{lccccccccl}
\toprule
 Hyperparameter & Type & Range \\
\midrule
Number of estimators & Categorical & [100, 200, 300, 1000]\\
Max depth & Categorical & [1,2,3,4,5,6]\\
Max features & Categorical & [2, 6]\\
Max samples & Categorical &[0.1, 0.3, 0.5, 0.7, 0.9]\\
\bottomrule
\end{tabular}
\end{table}

\subsection{Performance measures}
In addition to the above-mentioned AUROC, used to evaluate classifier performance in both the real and the simulated data sets, another performance measure is used for simulated data sets which benefits from the availability of the ground truth f-waves. Since the main distinction among extraction methods lies in their ability to cancel the QRS~complex, the RMS error is computed within the QRS interval, defined by 90~ms before and 90~ms after the R~peak. The RMS error is also computed outside the QRS~interval.

\subsection{Statistical analysis}
The f-wave features in windows classified as AF were analyzed by sex and age. The non-parametric Mann--Whitney rank test was used to determine whether the features were significantly different between men and women. $p < 0.05$ indicates a statistically significant difference. When evaluating features by age, the patients were divided into three age groups. The first group consisted of patients younger than 60 years old, the second group consisted of patients aged 60 or older but younger than 75, and the third group consisted of patients aged 75 or older.

%%%%%%%%%%%%%%%%%%%%%%%%%%%%%%%%%%%%%%%%%%%%%%%%%%%%%%%%%%%%%%%%%%%%%%%%%%%%
\section{Results}
\label{secResults}
\subsection{Extraction performance on real data sets}
Classifier performance is presented in Table~\ref{table:test_results} for each of the four extraction methods, each of the three data sets, and each of the two or three leads available. The results show that TS$_{\textrm{PCA}}$ yielded the largest AUROC, regardless of the data set and lead analyzed.% The ranking of the other three methods exploring the principle of average beat subtraction only vary slightly from set to set and lead to lead.

\begin{figure*}[t]
%\centering
\includegraphics[page=1,width=\textwidth]{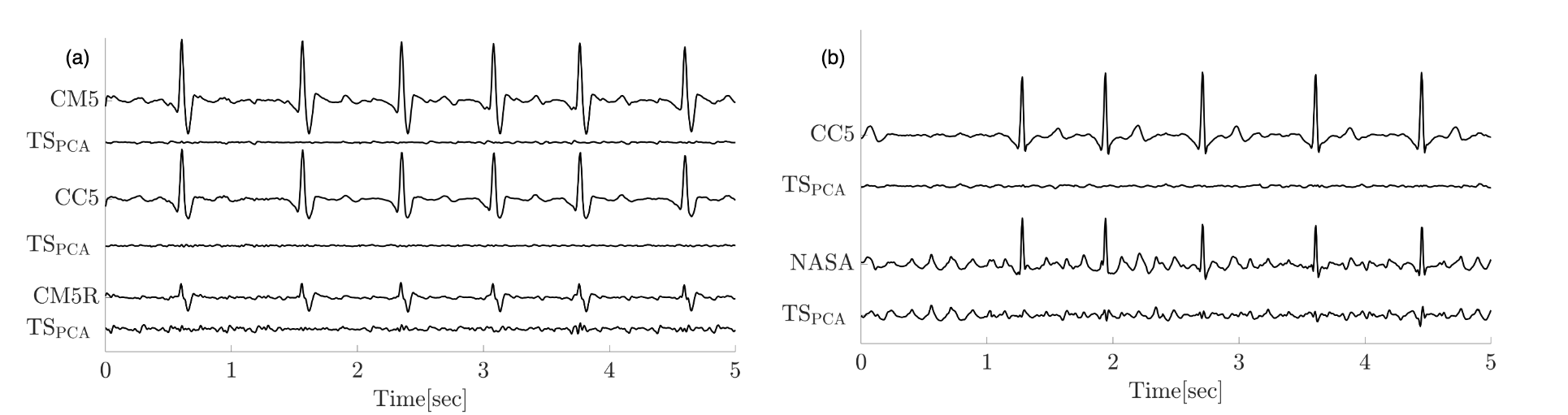}
\vspace{-0.4cm}
\caption{Illustration of f-wave extraction in ECGs excerpted from (a)~RBDB and (b)~SHDB, using~$\textrm{TS}_{\textrm{PCA}}$. }
\label{fig:af_example_different_leads}
\end{figure*}

Using TS$_{\textrm{PCA}}$, the largest AUROC is achieved for lead CM5R in RBDB and lead NASA in SHDB, i.e., 0.83 and 0.92, respectively, as these two leads are the ones nearest to the right atrium and with a vertical vector direction. Using TS$_{\textrm{PCA}}$, Fig.~\ref{fig:af_example_different_leads} illustrates f-wave extraction in different leads for ECGs excerpted from RBDB and SHDB. It should be noted that the lead with the largest AUROC for TS$_{\textrm{PCA}}$ is also the one with the largest AUROC for the other three methods. 
% Although the lead positions in UVAF are unknown, the results suggest that lead Ch3 corresponds to CM5R since the AUROC is the largest for Ch3.

%Although no quantification of ECG signal quality was performed in the present study, the much larger AUROC of SHDB suggests that the overall signal quality of this data set is considerably better, see Table~\ref{table:test_results}.

%%%%
\begin{table}
\caption{Performance of AF classification on the three real data test sets, expressed in terms of area under the receiver operating characteristic (AUROC). For each extraction method and each lead,the median AUROC results are presented, where the best results are indicated by boldface.}
\label{table:test_results}
\centering
\begin{adjustbox}{width=1\columnwidth,center=\columnwidth}
\small
\begin{tabular}{lccc|ccc|ccl}
\toprule
 &\multicolumn{3}{c|}{\textbf{UVAF}} & \multicolumn{3}{c|}{\textbf{RBDB}} & \multicolumn{2}{c}{\textbf{SHDB}}\\

 & \textbf{Ch1}& \textbf{Ch2} & \textbf{Ch3}  & \textbf{CM5}& \textbf{CC5} &  \textbf{CM5R} &\textbf{CC5}& \textbf{NASA}\\
\midrule
ABS &         0.69 & 0.75& 0.78 & 0.65 &0.60 & 0.81 & 0.86 & 0.87\\
ABS$_{\textrm{sc1}}$&  0.76 & 0.79& 0.81 & 0.62 &0.63 & 0.80 & 0.85 & 0.88\\
ABS$_{\textrm{sc2}}$ & 0.76 & 0.76& 0.81 & 0.63 &0.60 & 0.82 & 0.86 & 0.88\\
TS$_{\textrm{PCA}}$ & \textbf{0.81} &\textbf{0.83} & \textbf{0.85} & \textbf{0.72}& \textbf{0.67}& \textbf{0.83} & \textbf{0.87} & \textbf{0.92} \\
\bottomrule
\end{tabular}
\end{adjustbox}
\end{table}

%%%%

\subsection{Extraction performance on a simulated data set}
The AUROC obtained on the simulated data set are presented in Table~\ref{table:simulated_results} for each of the four extraction methods when  either noise-free or noisy ECGs are processed. While the four methods perform similarly on noise-free ECGs, the reduction in performance is more pronounced for the ABS-based methods than for $\textrm{TS}_{\textrm{PCA}}$ when noisy ECGS are processed.

Table~\ref{table:simulated_results2} presents the RMS error between  extracted and true f-waves, computed either inside or outside the QRS interval. The lowest RMS error inside the QRS interval is obtained for $\textrm{TS}_{\textrm{PCA}}$, being considerably lower than those of the ABS-based methods. As expected, the RMS error obtained outside the QRS interval is about the same for all four methods. Interestingly, $\textrm{TS}_{\textrm{PCA}}$ produces a lower RMS error inside than outside the QRS interval for both noise-free and moderately noisy ECGs---a result which stands in contrast to the ABS-based methods.

%got  results when comparing the whole extracted signal, inside the QRS interval and the outside the QRS. The results across the different methods are similar outside the QRS though still $\textrm{TS}_{\textrm{PCA}}$ is the lowest. The difference is the most significant inside the QRS interval.

\begin{table}
\caption{Performance of AF classification on the simulated data sets, expressed in terms of the median AUROC.}
\label{table:simulated_results}
\centering
\begin{tabular}{lcc}
\toprule
 &\multicolumn{2}{c}{\textbf{AUROC}} \\
noise level ($\mu$V RMS) & 0 & 100 \\
\midrule
ABS                   & 0.98 & 0.94 \\
ABS$_{\textrm{sc1}}$  & 0.97 & 0.93 \\
ABS$_{\textrm{sc2}}$  & 0.96 & 0.94 \\
TS$_{\textrm{PCA}}$   & \textbf{0.99} & \textbf{0.98} \\
\bottomrule
\end{tabular}
\end{table}
%ABS                   & 0.98 & 0.93 & 12.45 & 19.48 & 16.31 & 22.39 & 10.50  & 18.34\\
%ABS$_{\textrm{sc1}}$  & 0.97 & 0.92 & 11.16 & 18.45 & 12.28 & 18.68 & 10.45  & 18.29 \\
%ABS$_{\textrm{sc2}}$  & 0.96 & 0.94 & 11.26 & 18.28 & 13.05 & 18.77 & 10.31  & 18.01\\
%TS$_{\textrm{PCA}}$   & \textbf{0.99} & \textbf{0.98} & \textbf{9.83} & \textbf{17.39} & \textbf{8.30} & \textbf{15.86} & \textbf{10.10} & \textbf{17.77}\\

\begin{table}[t]
\caption{Performance of AF classification on the simulated data sets, expressed in terms of the RMS error between extracted and ground truth f-waves. The RMS error is computed inside and outside the QRS interval.}
\label{table:simulated_results2}
\centering
\begin{tabular}{lcccc}
\toprule
 & \multicolumn{4}{c}{\textbf{RMS error}} \\
 & \multicolumn{2}{c}{\textbf{inside QRS}} & \multicolumn{2}{c}{\textbf{outside QRS}}\\
noise level ($\mu$V RMS) & 0 & 100 & 0 & 100 \\
\midrule
ABS                      & 16 & 30 & 11  & 27\\
ABS$_{\textrm{sc1}}$     & 12 & 27 & 10  & 27 \\
ABS$_{\textrm{sc2}}$     & 13 & 27 & 10  & 27\\
TS$_{\textrm{PCA}}$      & \textbf{8} & \textbf{25} & 10 & \textbf{16}\\
\bottomrule
\end{tabular}
\end{table}

\subsection{Results across sex and age}
\begin{figure}[ht]
\centering
\includegraphics[page=1,width=1\columnwidth]{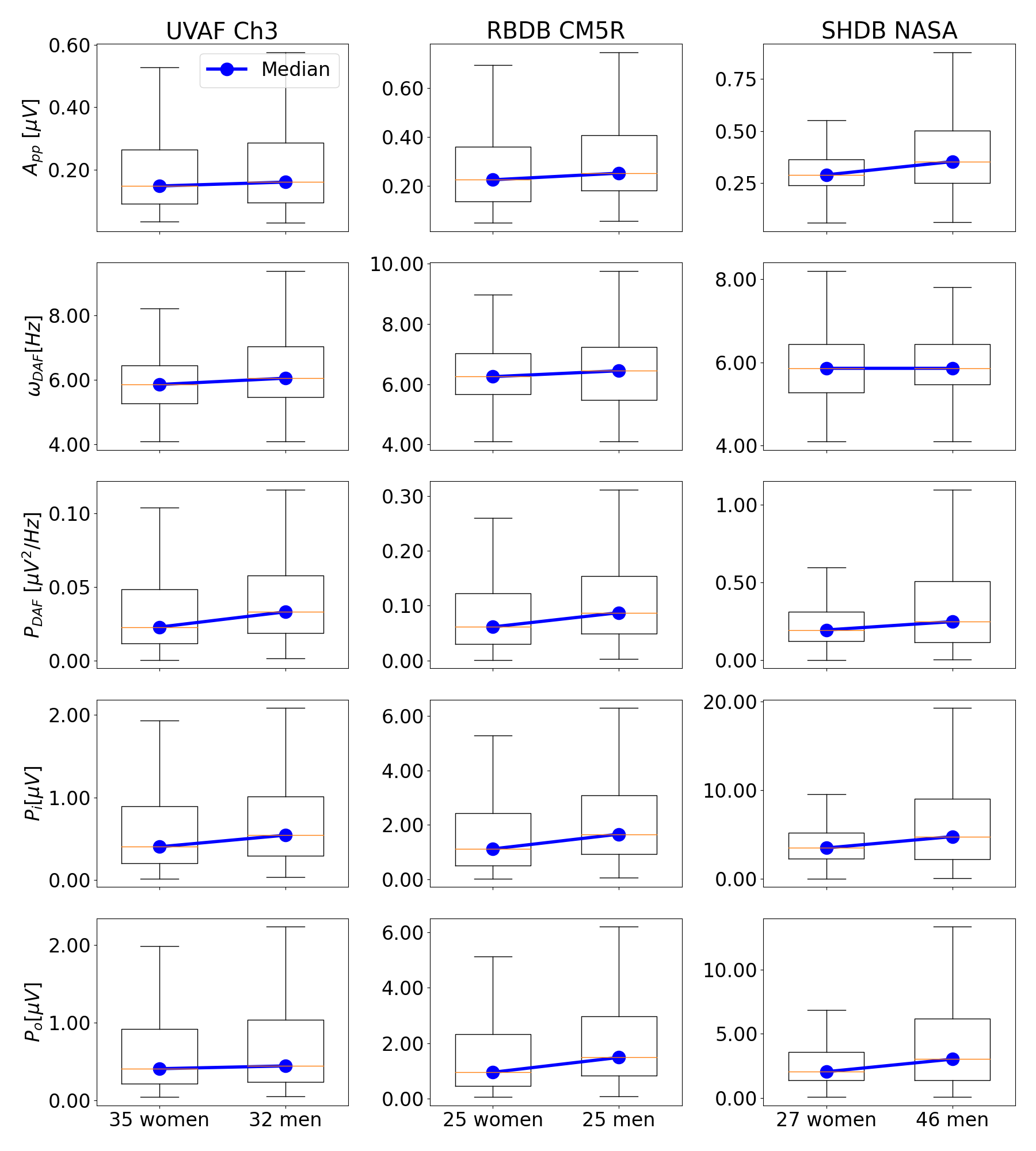}
\caption{Distributions of f-wave features as a function of sex, obtained on UVAF, RBDB, and SHDB. The first quartile, the median, and the third quartile are indicated by the box. The blue line is included to facilitate a comparison of women and men. The presented lead is selected based on the results presented in Table~\ref{table:test_results}. }
\label{sex_freq_feature}
\end{figure}

Figure~\ref{sex_freq_feature} presents the distributions of the five f-wave features as a function of sex, obtained on UVAF, RBDB, and SHDB. The results show that all features are significantly different between men and women. The medians of f-wave amplitude and spectral power of the DAF are lower in women than in men; this finding applies to all three data sets.

\begin{figure}[ht]
\centering
\includegraphics[page=1,width=1\columnwidth]{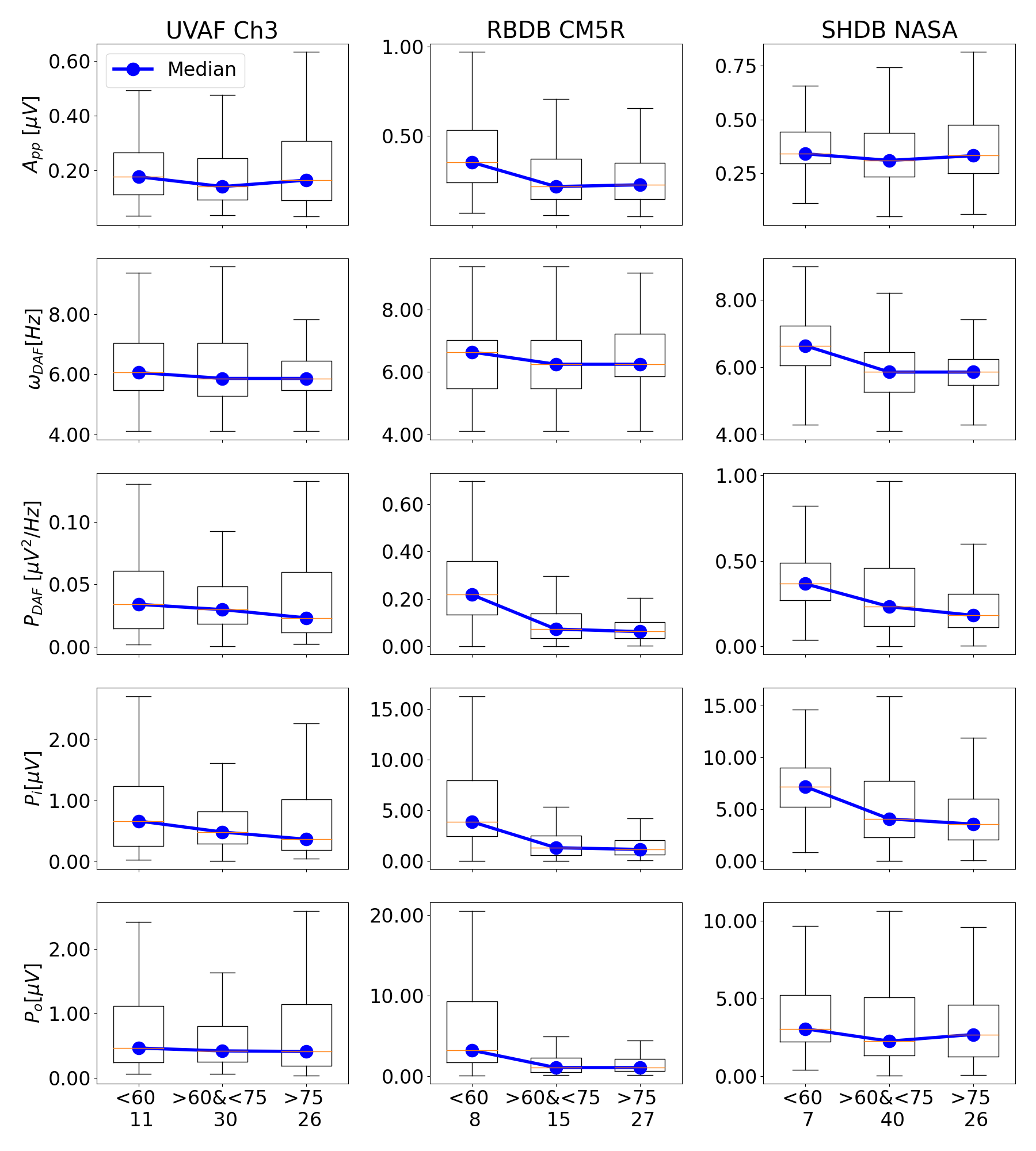}
\caption{Distributions of f-wave features as a function of age group, obtained on UVAF, RBDB, and SHDB, cf. Fig.~\ref{sex_freq_feature} for explanation of the displayed information. The number of patients in each age group is indicated at the bottom of the figure.}
%\caption{Age comparison: Box plot describing the distribution of the features in % UVAF, RBDB and SHDB between the different age groups. The blue line shows how %the median values across all data sets are highest in the youngest group of 
% patients that are younger than 60 years. These specific leads were selected %based on the results presented in Table \ref{table:test_results}.}
\label{age_freq_feature}
\end{figure}

Figure~\ref{age_freq_feature} presents the distributions of the five f-wave features as a function of the three age groups.  The results show that the youngest group, i.e. $<$60~years, exhibited higher medians compared to the other two groups.
%\footnote{For the sex comparison the median (Q1-Q3) values of the features are listed in Table~\ref{table:features_sex_comparison} and for the age comparison the values are listed in Table~\ref{table:features_age_comparison}.} 

%---------------------------------------------------------------------------------

\section{Discussion}  
\label{secDiscussion}
This study presents a novel approach to evaluating the performance of single-lead f-wave extraction methods, offering the distinct advantage of using large real data sets annotated with regard to the presence of AF. Another advantage is that the proposed approach does not measure performance in relation to the amount of QRS-related residuals as such measures run the risk of confusing a high noise level with large residuals. Yet another advantage is that the ranking of extraction methods is based on an AF classifier which takes into account the spectral properties of the f-waves.

% discussion of the significance of machine learning
The RF-based AF classifier was chosen because of its high accuracy and efficiency in handling large data sets. However, since the main purpose of the RF-based  classifier is to rank f-wave extraction methods, but not necessarily to offer the best AF classification performance, some other technique like a support vector machine or XGBoost may just as well be used instead. 

Of the four single-lead extraction methods evaluated, $\textrm{TS}_{\textrm{PCA}}$ consistently performed the best across the three real data sets and different lead positions. For comparison, the performance was evaluated on two simulated data sets, with or without noise, supporting the result on the real data sets  that $\textrm{TS}_{\textrm{PCA}}$  performed better than the three ABS-based methods, expressed in terms of the AUROC as well as the RMS error between extracted and ground truth f-waves. As for ranking of the ABS-based methods, the AUROC varies across the real data sets and lead positions as well as across the simulated data sets, suggesting that these three methods offer similar performance.  

In the present study, only a handful of well-known single-lead extraction methods were evaluated. However, several other single-lead extraction methods would deserve to be evaluated, based on, e.g., modeling of the QRST complex and the f-waves in combination with an extended  Kalman smoother~\cite{BaysianFiltering2007}, see also~\cite{Behar2014CombiningData}, resonance-based signal decomposition~\cite{Zhu2022F-WaveMethod}, and diffusion geometry data analysis expanding on the ABS principle~\cite{Malik2017Single-leadGeometry}.

%% discussion about lead position, sex and age.
The results vary with respect to lead position, showing that certain leads better capture the f-wave activity than others. Specifically, lead CM5R in RBDB and lead NASA in SHDB yielded the largest AUROC. Both these leads are roughly positioned along the same axis relative to the electrical vector of the heart and positioned close to the right atrium, thus explaining why the f-waves are better captured in these leads than in the other leads of the data sets~\cite{APetrenas2018LeadDevice,OptimumLeadPositioning}. Using the best-performing method, i.e., $\textrm{TS}_{\textrm{PCA}}$, and the lead associated with the largest AUROC of each real data set, and looking only at AF windows the distribution of the five f-wave features showed that men had higher median f-wave features values than women. It also showed that younger individuals had higher median f-wave features compared to older individuals.

%% perspective on using the f-wave in research and for clinical applications
Previous research has demonstrated that f-wave characteristics are associated with various clinical outcomes. For example, high f-wave amplitude is associated with a better prognosis in patients after AF ablation~\cite{highfwaves}. Low f-wave amplitude was shown to predict sinus node dysfunction following AF ablation in persistent AF \cite{Lowfwaves2015}. Fine f-waves,  defined by an amplitude less than 0.1~mV in all leads, is a risk factor for heart failure~\cite{kawaji2022association}. Evaluating the best f-wave extraction method using real ECGs can further strengthen research in cardiac electrophysiology by enabling a more precise f-wave characterization. For the clinical practice, single-lead f-wave extraction provides an opportunity to support the diagnosis of AF episodes by distinguishing AF from other supraventricular tachyarrhythmias and in those situations when AF is present but the heart rate is below 100~beats per minute. In addition, robust f-wave extraction could facilitate research on AF phenotypes that may be related to lead-dependent f-wave manifestations.

As already noted, a limitation of the present study is that only a handful of extraction methods were subject to evaluation. However, the main aim of the study was to develop an approach for ranking of methods, and, therefore, the number of methods was kept low. Another limitation is the exclusion of atrial flutter windows, representing an important supra\-ventricular tachyarrhythmia characterized by abnormal atrial activity, as well as mixed rhythms windows including both AF and non-AF beats. Since the F-waves in atrial flutter are more regular, the ranking of the extraction methods will most likely not be the same.  Yet another limitation is that the groups divided into different sex or age were small. Therefore, a future study with larger groups will provide stronger and more conclusive results regarding f-wave characteristics. Since medication can influence the DAF as well as other f-wave characteristics, see, e.g., ~\cite{Husser2007TFdrug}, medication should ideally be taken into account when evaluating f-wave characteristics. However, this information was only available in the RBDB and therefore not taken into consideration.

\section{Conclusions}
\label{secConclusion}
% We have developed a novel approach for ranking f-waves extraction methods for 1-lead systems. This method is capable of ranking extracted f-waves on real data without relying on ground truth f-waves. By using this method, researchers can identify the most suitable extraction method for their data and discover new characteristics related to f-waves in different age or sex groups.

This research offers a novel method for ranking of single-lead f-wave extraction methods based on AF classification performance. Four methods were evaluated on three real data sets from different geographical regions, where $\textrm{TS}_{\textrm{PCA}}$ was ranked as the best-performing method. Robust extraction would facilitate more advanced research on f-wave characteristics and their association with clinical outcomes. The source code of this work is made open source to facilitate further research (upon publication).

%The first main contribution of this research is the introduction of a novel method for assessing f-wave extraction performance. For that purpose four extraction methods were compared and evaluated on three different data sets from different geographical regions. $TS_{PCA}$ performed the best at extracting the f-wave (Table \ref{table:test_results}). This ranking was consistent across data sets from multiple centers as well as for the simulated data set. 

%\appendices

%\section*{Acknowledgment}
%This research was supported for NBM, SB and JB by a grant (3-17550) from the Ministry of Science \& Technology, Israel \& Ministry of Europe and Foreign Affairs (MEAE) and the Ministry of Higher Education, Research and Innovation (MESRI) of France. SB, NBM, MS and JAB acknowledge the support of the Technion-Rambam Initiative in Artificial Intelligence in Medicine, Hittman: Technion EVPR Fund: Hittman Family Fund and Israel PBC-VATAT and by the Technion Center for Machine Learning and Intelligent Systems (MLIS).

\printbibliography

@article{Mihandoost2022AExtraction,
	author = {Mihandoost, Sara and S{\"{o}}rnmo, Leif and Doyen, Matthieu and Oster, Julien},
	journal = {Physiol. Meas.},
	keywords = {ECG signal processing, atrial fibrillation, f-wave extraction},
	%pages = {105006},
	title = {{A comparative study of the performance of methods for f-wave extraction}},
	volume = {43},
	year = {2022}}

@article{Pan1985AAlgorithm,
	author = {Pan, Jiapu and Tompkins, Willis J.},
	journal = {IEEE. Trans. Biomed.},
%	number = {3},
	pages = {230--236},
	pmid = {3997178},
	title = {{A real-time QRS detection algorithm}},
	volume = {32},
	year = {1985}}

@article{Martens2007ARecordings,
	author = {Martens, Suzanna M M and Rabotti, Chiara and Mischi, Massimo and Sluijter, Rob J},
	journal = {Physiol. Meas.},
%	month = {4},
%	number = {4},
	pages = {373--388},
	title = {{A robust fetal ECG detection method for abdominal recordings}},
	volume = {28},
	year = {2007}}

@article{Alcaraz2008AdaptiveElectrocardiograms,
	author = {Alcaraz, Ral and Rieta, Jos{\'e} Joaqu{\'\i}n},
	journal = {Physiol. Meas.},
	pages = {1351--1369},
	pmid = {18946157},
	publisher = {Physiol Meas},
	title = {{Adaptive singular value cancelation of ventricular activity in single-lead atrial fibrillation electrocardiograms}},
	volume = {29},
	year = {2008}}

@article{Roonizi2017AnFibrillation,
	author = {Roonizi, Ebadollah Kheirati and Sassi, Roberto},
	journal = {IEEE J. Biomed. Health Inform.},
	keywords = {80 and over, Adult, Aged, Algorithms, Atrial Fibrillation / diagnosis*, Atrial Fibrillation / physiopathology*, Bayes Theorem, Computer-Assisted*, Ebadollah Kheirati Roonizi, Electrocardiography*, Female, Heart Atria / physiopathology*, Heart Ventricles / physiopathology*, Humans, MEDLINE, NCBI, NIH, NLM, National Center for Biotechnology Information, National Institutes of Health, National Library of Medicine, PubMed Abstract, Roberto Sassi, Signal Processing, Signal-To-Noise Ratio, Young Adult, doi:10.1109/JBHI.2016.2625338, pmid:27834661},
%	month = {11},
%	number = {6},
	pages = {1573--1580},
	pmid = {27834661},
	publisher = {IEEE J. Biomed. Health Inform.},
	title = {{An extended Bayesian framework for atrial and ventricular activity separation in atrial fibrillation}},
	volume = {21},
	year = {2017}}

@article{Rieta2004AtrialSeparation,
	author = {Rieta, Jos{\'e} Joaqu{\'\i}n and Castells, Francisco and S{\'{a}}nchez, C{\'e}sar and Zarzoso, Vicente and Millet, Jos{\'e}},
	journal = {IEEE. Trans. Biomed.},
	keywords = {Algorithms, Atrial Fibrillation / diagnosis*, Atrial Fibrillation / physiopathology*, Body Surface Potential Mapping / methods, Comparative Study, Computer-Assisted / methods*, Diagnosis, Electrocardiography / methods*, Evaluation Study, Francisco Castells, Heart Atria / physiopathology*, Heart Conduction System / physiopathology, Heart Ventricles / physiopathology, Humans, Jos{\'{e}} Joaqu{\'{i}}n Rieta, Jos{\'{e}} Millet, MEDLINE, NCBI, NIH, NLM, National Center for Biotechnology Information, National Institutes of Health, National Library of Medicine, Non-U.S. Gov't, Principal Component Analysis, PubMed Abstract, Reproducibility of Results, Research Support, Sensitivity and Specificity, Validation Study, doi:10.1109/TBME.2004.827272, pmid:15248534},
%	month = {7},
%	number = {7},
	pages = {1176--1186},
	pmid = {15248534},
	publisher = {IEEE Trans. Biomed. Eng.},
	title = {{Atrial activity extraction for atrial fibrillation analysis using blind source separation}},
	volume = {51},
	year = {2004}}

@article{Dai2013AtrialMethods,
	author = {Dai, Huhe and Jiang, Shouda and Li, Ye},
	journal = {Comput. Biol. Med.},
	keywords = {Algorithms*, Atrial Function / physiology*, Cardiovascular*, Electrocardiography / methods*, Heart Atria, Huhe Dai, Humans, MEDLINE, Models, NCBI, NIH, NLM, National Center for Biotechnology Information, National Institutes of Health, National Library of Medicine, PubMed Abstract, Shouda Jiang, Ye Li, doi:10.1016/j.compbiomed.2012.12.005, pmid:23318044},
%	month = {3},
%	number = {3},
	pages = {176--183},
	pmid = {23318044},
	publisher = {Comput Biol Med},
	title = {{Atrial activity extraction from single lead ECG recordings: evaluation of two novel methods}},
	volume = {43},
	year = {2013}}

@article{Biton2021AtrialLearning,
	author = {Biton, Shany and Gendelman, Sheina and Ribeiro, Ant{\^o}nio H and Miana, Gabriela and Moreira, Carla and Ribeiro, Antonio Luiz P and Behar, Joachim A},
	journal = {Eur. Heart J. -- Digital Health },
	volume = {2},
	publisher = {Oxford University Press (OUP)},
	title = {{Atrial fibrillation risk prediction from the 12-lead electrocardiogram using digital biomarkers and deep representation learning}},
    pages = {576–585},
	year = {2021}}

@article{Lemay2007CancellationMethods,
	author = {Lemay, Mathieu and Vesin, Jean Marc and Van Oosterom, Adriaan and Jacquemet, Vincent and Kappenberger, Lukas},
	journal = {IEEE. Trans. Biomed.},
	keywords = {Artifacts*, Atrial Fibrillation / diagnosis*, Atrial Fibrillation / physiopathology*, Comparative Study, Computer-Assisted / methods*, Diagnosis, Electrocardiography / methods*, Evaluation Study, Heart Conduction System / physiopathology*, Heart Ventricles / physiopathology*, Humans, Jean-Marc Vesin, Lukas Kappenberger, MEDLINE, Mathieu Lemay, NCBI, NIH, NLM, National Center for Biotechnology Information, National Institutes of Health, National Library of Medicine, Non-U.S. Gov't, PubMed Abstract, Reproducibility of Results, Research Support, Sensitivity and Specificity, doi:10.1109/TBME.2006.888835, pmid:17355069},
%	month = {3},
%	number = {3},
	pages = {542--546},
	pmid = {17355069},
	publisher = {IEEE Trans. Biomed. Eng.},
	title = {{Cancellation of ventricular activity in the ECG: Evaluation of novel and existing methods}},
	volume = {54},
	year = {2007}}

@article{Behar2014CombiningData,
	author = {Behar, Joachim A. and Oster, Julien and Clifford, Gari D.},
	journal = {Physiol. Meas.},
	keywords = {blind source separation, Kalman filter, non invasive foetal ECG, template substraction},
%	month = {8},
%	number = {8},
	pages = {1569--1589},
	pmid = {25069410},
	publisher = {IOP Publishing Ltd},
	title = {{Combining and benchmarking methods of foetal ECG extraction without maternal or scalp electrode data}},
	volume = {35},
	year = {2014}}

@article{Barrett2014ComparisonMonitoring,
	author = {Barrett, Paddy M. and Komatireddy, Ravi and Haaser, Sharon and Topol, Sarah and Sheard, Judith and Encinas, Jackie and Fought, Angela J. and Topol, Eric J.},
	journal = {Am. J. Med.},
	keywords = {Adhesives, Adult, Aged, Ambulatory*, Arrhythmias, Atrial Fibrillation / diagnosis, Atrial Flutter / diagnosis, Atrioventricular Block / diagnosis, Cardiac / diagnosis*, Cardiac / physiopathology, Comparative Study, Electrocardiography, Electrocardiography / methods*, Equipment Design, Eric J Topol, Extramural, Female, Humans, MEDLINE, Male, Middle Aged, N.I.H., NCBI, NIH, NLM, National Center for Biotechnology Information, National Institutes of Health, National Library of Medicine, PMC3882198, Paddy M Barrett, Patient Satisfaction, PubMed Abstract, Ravi Komatireddy, Research Support, Supraventricular / diagnosis, Tachycardia, Time Factors, Ventricular / diagnosis, Ventricular Fibrillation / diagnosis, doi:10.1016/j.amjmed.2013.10.003, pmid:24384108},
%	number = {1},
	pages = {11--95},
	pmid = {24384108},
	publisher = {Am J Med},
	title = {{Comparison of 24-hour Holter monitoring with 14-day novel adhesive patch electrocardiographic monitoring}},
	volume = {127},
	year = {2014}}

@incollection{Beckers2005DeterminationLeads,
	author = {Beckers, F and Ann{\'{e}}, W and Verheyden, B and Van Der Dussen De Kestergat, C and Van Herk, E and Janssens, L and Willems, R and Heidb{\"{u}}chel, H and Aubert, A E},
	booktitle = {Proc. Comput. Cardiol.},
	pages = {339--442},
	title = {{Determination of atrial fibrillation frequency using QRST-cancellation with QRS-scaling in standard electrocardiogram leads}},
	volume = {32},
	year = {2005}}

@article{Slocum1992DiagnosisActivity,
	author = {Slocum, Janet and Sahakian, Alan and Swiryn, Steven},
	journal = {J. Electrocardiol.},
	keywords = {A Sahakian, Algorithms*, Atrial Fibrillation / diagnosis*, Computer-Assisted*, Electrocardiography / methods*, Humans, J Slocum, MEDLINE, NCBI, NIH, NLM, National Center for Biotechnology Information, National Institutes of Health, National Library of Medicine, PubMed Abstract, S Swiryn, Sensitivity and Specificity, Signal Processing, doi:10.1016/0022-0736(92)90123-h, pmid:1735788},
%	number = {1},
	pages = {1--8},
	pmid = {1735788},
	publisher = {J Electrocardiol},
	title = {{Diagnosis of atrial fibrillation from surface electrocardiograms based on computer-detected atrial activity}},
	volume = {25},
	year = {1992}}

@article{Behar2013ECGReduction,
	author = {Behar, Joachim and Oster, Julien and Li, Qiao and Clifford, Gari D.},
	journal = {IEEE. Trans. Biomed.},
	keywords = {Electrocardiogram (ECG), intensive care unit (ICU), signal quality},
%	number = {6},
	pages = {1660--1666},
	pmid = {23335659},
	title = {{ECG signal quality during arrhythmia and its application to false alarm reduction}},
	volume = {60},
	year = {2013}}

@article{Steinhubl2018EffectTrial,
	author = {Steinhubl, Steven R. and Waalen, Jill and Edwards, Alison M. and Ariniello, Lauren M. and Mehta, Rajesh R. and Ebner, Gail S. and Carter, Chureen and Baca-Motes, Katie and Felicione, Elise and Sarich, Troy and Topol, Eric J.},
	journal = {JAMA},
	keywords = {atrial fibrillation, continuous ecg monitoring, electrocardiogram},
%	month = {7},
%	number = {2},
	pages = {146--155},
	pmid = {29998336},
	publisher = {American Medical Association},
	title = {{Effect of a home-based wearable continuous ECG monitoring patch on detection of undiagnosed atrial fibrillation: The mSToPS randomized clinical trial}},
	volume = {320},
	year = {2018}}

@article{APetrenas2017ElectrocardiogramEpisodes,
	author = {{A. Petr{\.{e}}nas} and {V. Marozas} and {A. Solo{\v{s}}enko} and {R. Kubilius} and {J. Skibarkien{\.{e}}3} and {J. Oster} and {L. S{\"{o}}rnmo}},
	journal = {Physiol. Meas.},
	pages = {2058--2080},
	title = {{Electrocardiogram modeling during paroxysmal atrial fibrillation: application to the detection of brief episodes}},
	volume = {38},
	year = {2017}}

@article{Castells2005EstimationConcepts,
	author = {Castells, Francisco and Mora, C. and Rieta, J. J. and Moratal-P{\'{e}}rez, D. and Millet, J.},
	journal = {Med. Biol. Eng. Comput.},
	keywords = {Algorithms, Ambulatory / instrumentation*, Atrial Fibrillation / diagnosis*, C Mora, Computer-Assisted, Electrocardiography, F Castells, Humans, J Millet, MEDLINE, NCBI, NIH, NLM, National Center for Biotechnology Information, National Institutes of Health, National Library of Medicine, Non-U.S. Gov't, Principal Component Analysis, PubMed Abstract, Research Support, Signal Processing, Validation Study, doi:10.1007/BF02351028, pmid:16411627},
%	month = {9},
%	number = {5},
	pages = {557--560},
	pmid = {16411627},
	publisher = {Med Biol Eng Comput},
	title = {{Estimation of atrial fibrillatory wave from single-lead atrial fibrillation electrocardiograms using principal component analysis concepts}},
	volume = {43},
	year = {2005}}

@article{Lee2012EventElectrocardiograms,
	author = {Lee, Jeon and Song, Mi Hye and Shin, Dong Gu and Lee, Kyoung Joung},
	journal = {Med. Biol. Eng. Comput.},
	keywords = {Algorithms*, Atrial Fibrillation / diagnosis*, Atrial Fibrillation / physiopathology*, Automated / methods*, Computer-Assisted / methods*, Diagnosis, Electrocardiography / methods*, Humans, Jeon Lee, Kyoung-joung Lee, MEDLINE, Mi-hye Song, NCBI, NIH, NLM, National Center for Biotechnology Information, National Institutes of Health, National Library of Medicine, Non-U.S. Gov't, Pattern Recognition, PubMed Abstract, Reproducibility of Results, Research Support, Sensitivity and Specificity, doi:10.1007/s11517-012-0931-7, pmid:22718318},
%	month = {8},
%	number = {8},
	pages = {801--811},
	pmid = {22718318},
	publisher = {Med Biol Eng Comput},
	title = {{Event synchronous adaptive filter based atrial activity estimation in single-lead atrial fibrillation electrocardiograms}},
	volume = {50},
	year = {2012}}

@incollection{LSornmo2018ExtractionWaves,
	author = {{L. S{\"{o}}rnmo} and {A. Petr{\.{e}}nas} and P. Laguna and {V. Marozas}},
	booktitle = {Atrial Fibrillation from an Engineering Perspective},
        city = {Lund},
        country = {Sweden},
	chapter = {5},
	editor = {{S{\"{o}}rnmo L}},
	pages = {137--220},
	publisher = {Springer Nature},
	title = {{Extraction of f waves}},
	year = {2018}}

@article{Zhu2022F-WaveMethod,
	author = {Zhu, Junjiang and Lv, Jintao and Kong, Dongdong},
	journal = {Entropy},
	keywords = {Atrial fibrillation, F-wave extraction, Genetic algorithm, Morphological component analysis, Resonance-based signal decomposition, Wavelet transform},
	%month = {Jun},
        %number = {6},
	publisher = {MDPI},
	title = {{f-wave extraction from single-lead electrocardiogram signals with atrial fibrillation by utilizing an optimized resonance-based signal decomposition method}},
	%pages = {812},
	volume = {24},
        note  = {{A}rt. no. 812},
	year = {2022}}

@article{Kanjilal1997FetalDecomposition,
	author = {Kanjilal, P P and Saha, Goutam},
	journal = {IEEE. Trans. Biomed.},
	pages = {51--59},
	title = {{Fetal ECG extraction from single-channel maternal ECG using singular value decomposition}},
	volume = {44},
	year = {1997}}

@article{Biton2023GeneralizableSexes,
	author = {Biton, Shany and Aldhafeeri, Mohsin and Marcusohn, Erez and Tsutsui, Kenta and Szwagier, Tom and Elias, Adi and Oster, Julien and Sellal, Jean Marc and Suleiman, Mahmoud and Behar, Joachim A},
	journal = {NPJ Digit. Med.},
	%pages = {44},
	title = {{Generalizable and robust deep learning algorithm for atrial fibrillation diagnosis across ethnicities, ages and sexes}},
	volume = {6},
	year = {2023}}

@incollection{APetrenas2018LeadDevice,
	author = {{A. Petr{\.{e}}nas} and {V. Marozas} and {L. S{\"{o}}rnmo}},
	booktitle = {Atrial Fibrillation from an Engineering Perspective},
	chapter = {2},
        city = {Lund},
        country = {Sweden},
	editor = {{L. S{\"{o}}rnmo}},
	pages = {25--48},
	publisher = {Springer Nature},
	title = {{Lead systems and recording devices}},
	year = {2018}}

@article{Moss2014LocalImplications,
	author = {Moss, Travis J and Lake, Douglas E and Moorman, J Randall},
	journal = {Physiol. Meas.},
%	month = {10},
%	number = {10},
	pages = {1929--1942},
	title = {{Local dynamics of heart rate: detection and prognostic implications}},
	volume = {35},
	year = {2014}}

@article{Mateo2013RadialFibrillation,
	author = {Mateo, Jorge and Rieta, Jos{\'e} Joaqu{\'{i}}n},
	journal = {Comput. Biol. Med.},
	keywords = {Algorithms*, Atrial Fibrillation / physiopathology*, Computer Simulation, Computer*, Computer-Assisted*, Databases, Electrocardiography / instrumentation, Electrocardiography / methods*, Factual, Humans, Jorge Mateo, Jos{\'{e}} Joaqu{\'{i}}n Rieta, MEDLINE, NCBI, NIH, NLM, National Center for Biotechnology Information, National Institutes of Health, National Library of Medicine, Neural Networks, Non-U.S. Gov't, PubMed Abstract, Research Support, Signal Processing, doi:10.1016/j.compbiomed.2012.11.007, pmid:23228480},
%	month = {2},
%	number = {2},
	pages = {154--163},
	pmid = {23228480},
	publisher = {Comput Biol Med},
	title = {{Radial basis function neural networks applied to efficient QRST cancellation in atrial fibrillation}},
	volume = {43},
	year = {2013}}

@article{Kligfield2007RecommendationsElectrocardiogram,
	author = {Kligfield, Paul and Gettes, Leonard S. and Bailey, James J. and Childers, Rory and Deal, Barbara J. and Hancock, E. William and Van Herpen, Gerard and Kors, Jan A. and Macfarlane, Peter and Mirvis, David M. and Pahlm, Olle and Rautaharju, Pentti and Wagner, Galen S.},
	journal = {Circulation},
	keywords = {AHA Scientific Statements, computers, diagnosis, electrocardiography, electrophysiology, intervals, potentials, tests},
%	month = {3},
%	number = {10},
	pages = {1306--1324},
	pmid = {17322457},
	publisher = {Lippincott Williams {\&} Wilkins},
	title = {{Recommendations for the standardization and interpretation of the electrocardiogram}},
	volume = {115},
	year = {2007}}

@article{Chocron2021RemoteNetwork,
	arxivid = {2008.02228},
	author = {Chocron, Armand and Oster, Julien and Biton, Shany and Mandel, Franck and Elbaz, Meyer and Zeevi, Yehoshua Y. and Behar, Joachim A.},
	journal = {IEEE. Trans. Biomed.},
	keywords = {Armand Chocron, Atrial Fibrillation* / diagnosis, Computer, Databases, Electrocardiography, Entropy, Factual, Humans, Joachim A Behar, Julien Oster, MEDLINE, NCBI, NIH, NLM, National Center for Biotechnology Information, National Institutes of Health, National Library of Medicine, Neural Networks, PubMed Abstract, doi:10.1109/TBME.2020.3042646, pmid:33275575},
%	month = {8},
%	number = {8},
	pages = {2447--2455},
	pmid = {33275575},
	publisher = {IEEE. Trans. Biomed.},
	title = {{Remote atrial fibrillation burden estimation using deep recurrent neural network}},
	volume = {68},
	year = {2021}}

@article{Li2008RobustSources,
	author = {Li, Q and {Mark RG} and {Clifford GD}},
	journal = {Physiol Meas.},
	pages = {15--32},
	title = {{Robust heart rate estimation from multiple asynchronous noisy sources}},
	volume = {29},
	year = {2008}}

@article{Svennberg2017SafeFibrillation,
	author = {Svennberg, Emma and Stridh, Martin and Engdahl, Johan and Al-Khalili, Faris and Friberg, Leif and Frykman, Viveka and Rosenqvist, M{\aa}rten},
	journal = {EP Europace},
	keywords = {Algorithms, Atrial fibrillation, Electrocardiography, Mass screening, Stroke},
%	month = {9},
%	number = {9},
	pages = {1449--1453},
	pmid = {28339578},
	publisher = {Oxford Academic},
	title = {{Safe automatic one-lead electrocardiogram analysis in screening for atrial fibrillation}},
	volume = {19},
	year = {2017}}

@article{Malik2017Single-leadGeometry,
	author = {Malik, John and Reed, Neil and Wang, Chun-Li and Wu, Hau-Tieng},
	journal = {Physiol. Meas},
	pages = {1310--1334},
	title = {{Single-lead f-wave extraction using diffusion geometry}},
	volume = {38},
	year = {2017}}

@article{Stridh2001SpatiotemporalFibrillation,
	author = {Stridh, M. and S{\"{o}}rnmo, L.},
	journal = {IEEE. Trans. Biomed.},
	keywords = {Atrial Fibrillation / diagnosis*, Cardiovascular, Comparative Study, Computer-Assisted*, Humans, L S{\"{o}}rnmo, M Stridh, MEDLINE, Models, NCBI, NIH, NLM, National Center for Biotechnology Information, National Institutes of Health, National Library of Medicine, Non-U.S. Gov't, PubMed Abstract, Reference Values, Research Support, Signal Processing, Vectorcardiography*, doi:10.1109/10.900266, pmid:11235581},
%	number = {1},
	pages = {105--111},
	pmid = {11235581},
	publisher = {IEEE. Trans. Biomed.},
	title = {{Spatiotemporal QRST cancellation techniques for analysis of atrial fibrillation}},
	volume = {48},
	year = {2001}}

@article{Bataillou1995WeightedWeights,
	author = {Bataillou, E. and Thierry, E. and Rix, H. and Meste, O.},
	journal = {Signal Process.},
	keywords = {Adaptive process, Constrained minimization, ECG signal, Weighted averaging},
%	month = {6},
%	number = {1},
	pages = {51--66},
	publisher = {Elsevier},
	title = {{Weighted averaging using adaptive estimation of the weights}},
	volume = {44},
	year = {1995}}

@article{Chugh2014WorldwideStudy,
	author = {Chugh, Sumeet S. and Havmoeller, Rasmus and Narayanan, Kumar and Singh, David and Rienstra, Michiel and Benjamin, Emelia J. and Gillum, Richard F. and Kim, Young Hoon and McAnulty, John H. and Zheng, Zhi Jie and Forouzanfar, Mohammad H. and Naghavi, Mohsen and Mensah, George A. and Ezzati, Majid and Murray, Christopher J.L.},
	journal = {Circulation},
	keywords = {Adult, Age Distribution, Aged, Atrial Fibrillation / epidemiology*, Atrial Fibrillation / mortality*, Christopher J L Murray, Cost of Illness*, Extramural, Female, Global Health / statistics {\&} numerical data*, Humans, Incidence, MEDLINE, Male, Middle Aged, N.I.H., NCBI, NIH, NLM, National Center for Biotechnology Information, National Institutes of Health, National Library of Medicine, Non-U.S. Gov't, PMC4151302, Prevalence, PubMed Abstract, Quality-Adjusted Life Years, Rasmus Havmoeller, Research Support, Risk Factors, Sex Distribution, Sumeet S Chugh, doi:10.1161/CIRCULATIONAHA.113.005119, pmid:24345399},
%	month = {2},
%	number = {8},
	pages = {837--847},
	pmid = {24345399},
	publisher = {Circulation},
	title = {{Worldwide epidemiology of atrial fibrillation: A Global Burden of Disease 2010 Study}},
	volume = {129},
	year = {2014}}

@article{Kaasenbrood2016YieldVaccination,
	author = {Kaasenbrood, Femke and Hollander, Monika and Rutten, Frans H. and Gerhards, Leo J. and Hoes, Arno W. and Tieleman, Robert G.},
	journal = {Europace},
	keywords = {80 and over, Adult, Age Distribution, Aged, Atrial Fibrillation / diagnosis*, Atrial Fibrillation / epidemiology*, Electrocardiography / instrumentation*, Female, Femke Kaasenbrood, Humans, Influenza Vaccines / administration {\&} dosage, Logistic Models, MEDLINE, Male, Mass Screening / methods*, Middle Aged, Monika Hollander, Multivariate Analysis, NCBI, NIH, NLM, National Center for Biotechnology Information, National Institutes of Health, National Library of Medicine, Netherlands / epidemiology, PMC5072135, Primary Health Care, PubMed Abstract, Robert G Tieleman, Sex Distribution, Stroke / complications, doi:10.1093/europace/euv426, pmid:26851813},
%	month = {10},
%	number = {10},
	pages = {1514--1520},
	pmid = {26851813},
	publisher = {Europace},
	title = {{Yield of screening for atrial fibrillation in primary care with a hand-held, single-lead electrocardiogram device during influenza vaccination}},
	volume = {18},
	year = {2016}}

@article{Behar_2016,
year = {2016},
%month = {apr},
publisher = {IOP Publishing},
volume = {37},
%number = {5},
pages = {R1-R35},
author = {Behar, J and Andreotti, F and  Zaunseder, S and Oster, J and  Clifford, GD},
title = {A practical guide to non-invasive foetal electrocardiogram extraction and analysis},
journal = {Physiol. Meas.}}

@incollection{BenMoshe2022arnetECG,
  title={ArNet-{ECG}: Deep Learning for the detection of atrial fibrillation from the raw electrocardiogram},
  author={Ben-Moshe, N. and Biton, S. and Behar, J. A},
  booktitle = {Proc. Comput. Cardiol.},
  year={2022},
  volume ={49},
  pages = {1-4},
  city={Tampere},
  state={Finland}
}

@article{OptimumLeadPositioning,
year={1998},
%month={nov},
author={Waktare, J E  and Gallagher, M M and Murtagh, A and Camm, A J and Malik, M},
title= {Optimum lead positioning for recording bipolar atrial electrocardiograms during sinus rhythm and atrial fibrillation},
journal={Clin. Cardiol.},
pages={825-830},
volume={21}
}

@article{highfwaves,
year={2023},
%month={Jan},
author={Ishihara, S  and Maruyama, M and Nohara, T and Shimizu, W and Asai, K. },
title= {Atrial fibrillatory wave amplitude revisited: A predictor of recurrence after catheter ablation independent of the degree of left atrial structural remodeling},
journal={Cardiol J.},

}

@article{kawaji2022association,
  title={Association of inverted {T} wave during atrial fibrillation rhythm with subsequent cardiac events},
  author={Kawaji, Tetsuma and Ogawa, Hisashi and Hamatani, Yasuhiro and Kato, Masashi and Yokomatsu, Takafumi and Miki, Shinji and Abe, Mitsuru and Akao, Masaharu},
  journal={Heart},
  volume={108},
%  number={3},
  pages={178--185},
  year={2022},
  publisher={BMJ Publishing Group Ltd and British Cardiovascular Society}
}

@article{Lowfwaves2015,
  title={A low fibrillatory wave amplitude predicts sinus node dysfunction after catheter ablation in patients with persistent atrial fibrillation.},
  author={Sunaga, A. and  Masuda, M. and Kanda, T. and  Fujita, M. and Iida, O. and Okamoto, S. and Ishihara, T. and  Matsuda, Y.and Watanabe, T. and Sakata, Y. and Uematsu, M.},
  journal={J. Interv. Card.},
  volume={43},
  pages={253-261},
  year={2015},

}

@article{BaysianFiltering2007,
  title={A nonlinear {B}ayesian filtering framework for {ECG} denoising.},
  author={Sameni, R and Shamsollahi, M B and Jutten, C and Clifford, G D},
  journal={IEEE. Trans. Biomed.},
  volume={12},
  pages={2172-2185},
  year={2007},
%  month={Nov},
}

@article{Husser2007TFdrug,
	author = {D. Husser and M. Stridh and D. S. Cannom and A. K. Bhandari and M. J. Girsky and S. Kang and L. S{\"o}rnmo and S. B. Olsson and A. Bollmann},
	journal = {J. Cardiovasc. Electrophysiol.},
	pages = {41-46},
	title = {Validation and clinical application of time--frequency analysis of atrial fibrillation electrocardiograms},
	volume = {18},
	year = {2007}}

 \newpage
\supplementarysection
\begin{figure*}[t]
\centering
\includegraphics[page=1,width=1\textwidth]{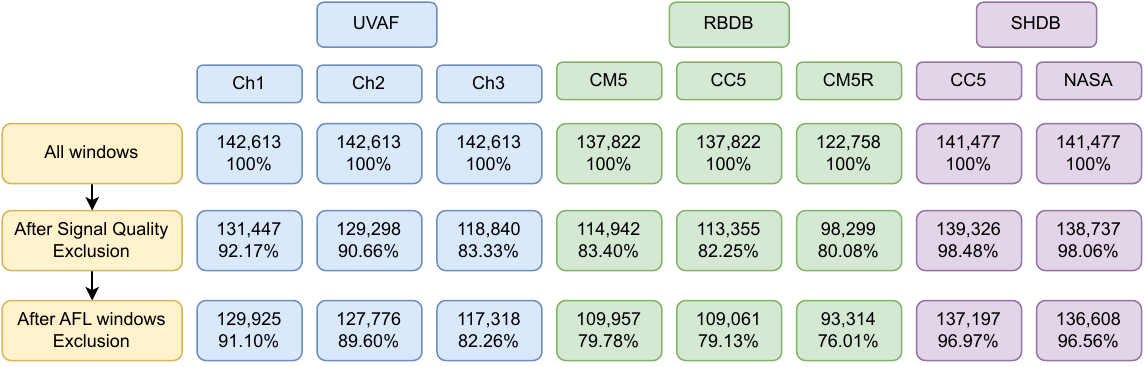}
\caption{Exclusion criteria and number of windows excluded in each lead of each data set.}
\label{exclusion_of_windows}
\end{figure*}

\end{document}